\documentstyle[twoside,fleqn,espcrc2,epsfig]{article}

\newcommand{\AmS}{{\protect\the\textfont2
  A\kern-.1667em\lower.5ex\hbox{M}\kern-.125emS}}
\def\beq{\begin{equation}}
\def\eeq{\end{equation}}
\def\bea{\begin{eqnarray}}
\def\eea{\end{eqnarray}}
\def\bq{\begin{quote}}
\def\eq{\end{quote}}

\def\nnb{\nonumber}
\def\ga{\left(}
\def\dr{\right)}
\def\aga{\left\{}
\def\adr{\right\}}

\def\rar{\rightarrow}

\def\nnb{\nonumber}
\def\la{\langle}
\def\ra{\rangle}
\def\nin{\noindent}
\def\ba{\begin{array}}
\def\ea{\end{array}}
\def\bm{\overline{m}}

\def\als{\alpha_s}
\def\as{\ga\frac{\bar{\alpha_s}}{\pi}\dr}
\def\asr{\ga\frac{{\alpha_s}}{\pi}\dr}

\title{\bf{Extracting $\overline{m}_c(M_c)$ and 
$f_{D_s,B}$ from the pseudoscalar sum rules}}
\author{ Stephan Narison\address{
Laboratoire de Physique Math\'ematique,
Universit\'e de Montpellier II
Place Eug\`ene Bataillon,
34095 - Montpellier Cedex 05, France.\\
E-mail:
narison@lpm.univ-montp2.fr}
\thanks{Talk given at the QCD 98 Euroconference-Montpellier
(2-8th July1998) -- Montpellier preprint PM-98/36. }}
\begin{document}
\pagestyle{plain}
\begin{abstract}
\noindent
I report here on the (first) direct extraction of the running
charm quark mass $\bar m_c(\nu)$ from the $D$-meson sum rules, and on the implications of
this result for the estimate of the leptonic decay constants $f_{D_{(s)}}$. 
The outputs: $\bar m_c(M_c)=(1.08\pm 0.11)$ GeV, 
$f_{D}\simeq (1.52\pm 0.16)f_\pi$, 
$f_{D_s}\simeq (1.75\pm 0.18)f_\pi$ and $f_B=(1.44\pm 0.07)f_\pi$ are in good agreement
with the existing sum rule results obtained using the pole mass. In particular, the result
$f_D\approx f_B$ supports early '87 sum rule results \cite{SNFB}, which indicated
a huge $1/m$ correction to the heavy quark symmetry expectation. This talk is based on 
the paper hep-ph/9712386 and updates the discussions given there.
\end{abstract}
\maketitle
\section{Introduction}
One of the most important parameters of the standard model is the
quark masses. However, contrary to the leptons, where the physical mass
can be identified with the pole of the propagator, the quark masses are
difficult to define because of confinement. Some attempts have been
made in order to define the quark pole mass within perturbation theory,
where it has been shown to be IR-finite \cite{TARRACH} and independent
from the choice of the regularization and renormalization schemes used \cite{SNPOLE}.
More recently, it has been noticed, in the limit of a large number of flavours,
that the resummation of perturbative series can induce a non-perturbative term,
which can affect the truncated perturbative result, 
and can, then, limit the accuracy of
the pole mass determination \cite{BENEKE}.  However, a proper use of such a result 
needs a resummation, at the same
level of accuracy (which is not often the case), of the Green function of a given
process involving the pole mass, where some eventual cancellation
between the resummed terms of the two series may occur (we plan to come back to this
point in a future publication). One may bypass the previous problems, by working, at a
given order of perturbative QCD, with the running quark masses, which are treated like
coupling constants of the QCD Lagrangian (see e.g. \cite{FNR}), and where some 
non-perturbative-like effect is expected to be absent. 
A lot of efforts has been furnished in the
literature
\cite{PDG} for extracting directly from the data the running masses of the light
 and heavy quarks 
using the SVZ QCD spectral sum rules (QSSR) \cite{SVZ,SNB}. In this note, I
shall consider a direct extraction of the running charm quark mass
using the observed value of $M_D=1.865$ MeV, and study its implication
on the value of the decay constants
$f_{D_{(s)}}$, which are normalized as $f_\pi=93.3$ MeV
 where the leptonic decay width reads:
\bea
\Gamma(D_s\rar l\nu(\gamma))&=&\frac{G^2_F|V_{cs}|^2}{4\pi}\nnb\\
&\times&f^2_{D_s}m^2_lM_{D_s}\ga
1-\frac{m_l^2}{M^2_B}\dr^2 \nnb\\
\eea
In this respect, this present work is an improvement and update of the ones in
\cite{SNFD,SNFB} based on the use of the perturbative pole mass, where some eventual
non-perturbative effects induced by the resummation of the QCD series are not taken
into account. 
For our discussion, we shall use the average value of the experimental widths quoted in
\cite{HEP96}, from which we can deduce:
\beq\label{fdsex}
f_{D_s}\simeq (1.92\pm 0.23)f_\pi.
\eeq
while, combining it with the most reliable sum rule result for the ratio
\cite{SNFD}:
\beq\label{fdratio}
{f_{D_s}}/{f_D}=1.15\pm 0.04~
\eeq
one also obtains:
\beq\label{fdex}
f_{D}\simeq (1.67\pm 0.24)f_\pi.
\eeq
\section{The QCD spectral sum rules}
 We shall work with the pseudoscalar
two-point correlator: 
\beq
\psi_5(q^2) \equiv i \int d^4x ~e^{iqx} \
\la 0\vert {\cal T}
J_q(x)
J^\dagger _q(0) \vert 0 \ra ,
\eeq
built from the heavy-light quark current:
$
J_d(x)=(m_c+m_d)\bar c(i\gamma_5)d,
$
and which has the quantum numbers of the $D$ meson.
The corresponding Laplace transform sum rules are:
\bea
{\cal L}(\tau)
&=& \int_{t_\leq}^{\infty} {dt}~\mbox{e}^{-t\tau}
~\frac{1}{\pi}~\mbox{Im} \psi_5(t),~~~{\mbox {and}}~~~\nnb\\
{\cal R}(\tau) &\equiv& -\frac{d}{d\tau} \log {{\cal L}(\tau)},
\eea
where $t_\leq$ is the hadronic threshold. The latter sum  rule,
 or its slight modification, is also useful, as it is equal to the 
resonance mass squared, in  
 the simple duality ansatz parametrization of the spectral function:
\bea
\frac{1}{\pi}\mbox{ Im}\psi_5(t)\simeq&& 2f^2_DM_D^4\delta(t-M^2_D)
 \ + \ \nnb\\
 &&``\mbox{QCD continuum}" \Theta (t-t_c),
\eea
where the ``QCD continuum comes from the discontinuity of the QCD
diagrams, which is expected to give a good smearing of the
different radial excitations \footnote{At
the optimization scale, its effect is negligible, such that a more
involved parametrization is not necessary.}. The decay constant $f_D$ is
analogous to $f_\pi=93.3$ MeV; 
$t_c$ is the QCD continuum threshold, which is, like the 
sum rule variable $\tau$, an  a priori arbitrary 
parameter. In this
paper, we shall impose the
 $t_c$ and $\tau$ stability criteria for extracting our optimal
results \footnote{The corresponding $t_c$ value very roughly indicates
the position of the next radial excitations.}. 
The QCD expression of the correlator
is known to two-loop accuracy
(see e.g. \cite{SNB} and the explicit expressions given in \cite{SNFB}),
in terms  of the perturbative pole mass $M_c$, and including the non-perturbative
condensates of dimensions less than or equal to six
\footnote{We shall use the corrected coefficient of the quark-gluon mixed
condensate given in \cite{SOTTO}. This change affects only slightly the result. We shall
also skip the negligible contribution from the dimension six four-quark 
and three-gluon condensates. Notice that
there is some discrepancy on the value of the four-quark coefficient in the literature.}. The sum rule
reads:
\bea
{\cal L}(\tau)
&=& M^2_c\Bigg{\{}\int_{4M^2_c}^{\infty} {dt}~\mbox{e}^{-t\tau}~\frac{3}{8\pi^2}
t(1-x)^2\Big{[} 1\nnb\\
&+&\frac{4}{3}\asr f(x)\Big{]} \nnb\\
&+&\Big{[} C_4\la O_4\ra +C_6\la
O_6\ra\tau\Big{]}~\mbox{e}^{-M^2_c\tau}\Bigg{\}},
\eea
with:
\bea
x&\equiv& M^2_c/t,\nnb\\
f(x)&=&\frac{9}{4}+2\rm{Li}_2(x)+\log x \log (1-x)\nnb\\&-&\frac{3}{2}\log (1/x-1)
-\log (1-x)\nnb\\ &+& x\log (1/x-1)-(x/(1-x))\log x, \nnb\\
C_4\la O_4\ra&=&-M_c\la \bar dd\ra -\la \als G^2\ra/12\pi\nnb\\
C_6\la O_6\ra&=&\frac{M^3_c\tau}{2}
g\la\bar d\sigma_{\mu\nu}\frac{\lambda_a}{2}G_a^{\mu\nu}d\ra
\eea 
It can be expressed in terms of the running mass $\bar{m}_c(\nu)$
\footnote{It is clear that, for the non-perturbative terms which are known to leading order
of perturbation theory, one can use either the running or the pole mass. However, we shall see
that the non-perturbative effects are not important in the analysis, such that this distinction
does not affect the result.},
 through
the perturbative  two-loop relation \cite{SNPOLE}:
\bea\label{relation}
M_c(\nu)=\bar{m}_c(\nu)\aga 1+
\as\ga \frac{4}{3}+2\log{\frac{\nu}{M_c}}\dr
\adr,
\eea
with: \cite{FNR,SNB}:
\bea
\bm_i(\nu)&=&\hat{m}_i\ga -\beta_1 \as(\nu)\dr^{-\gamma_1/\beta_1}\Bigg\{1+\nnb\\
&&\frac{\beta_2}{\beta_1}\ga \frac{\gamma_1}{\beta_1}-
 \frac{\gamma_2}{\beta_2}\dr\as(\nu)~+...\Bigg\},
\eea
where 
$\gamma_m$ and $\beta$ are respectively the QCD $\beta$-function
and mass anomalous dimension normalized as:
\beq
\beta(\als)=\sum_{i=1}^n\beta_i\as^i, ~~
\gamma_m=\sum_{i=1}^n\gamma_i\as^i,
\eeq
which, read for $n_f$ flavours \cite{SNB}:
\bea
\beta_1&=&-\frac{1}{2}\ga 11-\frac{2}{3}n_f\dr~,\nnb\\
\beta_2&=&-\frac{1}{4}\ga 51-\frac{19}{3}n_f\dr~,\nnb\\
\gamma_1&=&2~,~~~~~~~~~~\gamma_2=\frac{1}{6}\ga
\frac{101}{2}-\frac{5}{3}n_f\dr~.
\eea
As discussed earlier, non-perturbative terms induced by the resummation 
of the perturbative series can affect this relation \cite{BENEKE}. However, within
the approximation at which the spectral function is given, the use of this
perturbative relation should provide the correct expression of the spectral
function in terms of the running quark mass. In this way, unlike the analysis in
\cite{SNFD,SNFB}, the one done in this paper is not affected by the eventual
 existence of such non-perturbative terms related to the 
implicit use of the pole mass in the previous sum rule analysis. One should also
notice that, to the order we are working, the expression of the spectral function
in terms of the running and pole masses differ in the $\alpha_s$ correction, 
induced by the overall leading $M^2_c$ term appearing in Eq. (6). 
Throughout this paper we
shall use the values of the parameters
\cite{SNB,SNG}  given in Table 1. We have used for the mixed condensate the
parametrization:
\bea
g\la\bar d\sigma_{\mu\nu}\frac{\lambda_a}{2}G_a^{\mu\nu}d\ra&=&M^2_0\la\bar dd\ra,
\eea
where $M^2_0=(.8\pm .1)$ GeV$^2$ \cite{SNB}.
We shall also use, for four active flavours \cite{PDG}:
\beq
\Lambda= (325\pm 100)~{\mbox{MeV}}.
\eeq
One can also inspect that the dominant non-perturbative contribution  is due to the
dimension-four $m_c\la \bar dd\ra$ light quark condensate, while the other
non-perturbative effects remain a small correction
at the optimization scale, which corresponds to $\tau\simeq 0.6\sim 1$ GeV$^{-2}$ and $t_c
\simeq 6\sim 8$ GeV$^2$.
\section{Discussions and results }
\begin{figure}[hbt]
\begin{center}
\epsfxsize=7.5cm\epsfbox{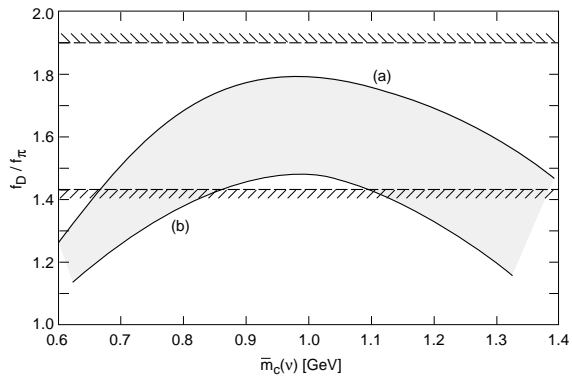}
\caption{Behaviour of $f_D$ versus $\bar m_c(\nu)$. The
horizontal band is the experimental domain of $f_D$. The theoretical band is limited by the
two curves: (a) $\Lambda$=0.425 GeV, $\nu$=1 GeV $\approx \tau^{-1/2}$, $t_c\geq 7$ GeV$^{2}$,
$\la \bar dd\ra^{1/3}$(1 GeV)=$-238$ MeV, 
$M^2_0=0.9$ GeV$^2$, 
$\la \alpha_s G^2\ra=0.06$ GeV$^2$
 and (b)
$\Lambda$=0.225 GeV, $\nu\approx M_c=1.42~\rm{GeV}$, $t_c=6$ GeV$^{2}$,
$\la \bar dd\ra^{1/3}$(1 GeV)=$-220$ MeV, 
$M^2_0=0.6$ GeV$^2$, 
$\la \alpha_s G^2\ra=0.08$ GeV$^2$.}
\end{center}
\end{figure}
\nin
Given the experimental value
on $M_D=1.865$ GeV, we present our results on $f_D$ from the first sum rule
for different values of the charm quark running mass evaluated at
$p^2=\nu^2$ in Fig. 1. The second sum rule gives the prediction on $M_D$ for
each   value of the charm mass (Fig. 2). The theoretical band is limited by the two extremal
values of the QCD parameters used in Table 1. Notice that the effect of the errors of the
different input parameters is much smaller in the ratio of sum rule ${\cal R}(\tau)$. The previous
analysis leads to the prediction from a two-loop calculation in the $\overline{MS}$ scheme:
\beq
\bar m_c(M_c)=(1.08\pm 0.11)~\mbox{GeV},
\eeq
\begin{table}[hbt]
\setlength{\tabcolsep}{0.5pc}
\caption{Different sources of errors in the estimate of $f_D$}
\begin{tabular}{c c }
\hline 
 & \\
Sources&$|\Delta (f_D/f_\pi)|$\\
&\\
\hline
&\\
$\Lambda=(325\pm 100)$ MeV&0.12\\
$\nu=(1.20\pm 0.22)$ GeV&0.08\\
$\bar m_c(M_c)=(1.20\pm 0.05)$ GeV&0.05\\
$t_c=(6.5\pm 0.5)$ GeV$^2$&0.04\\
$\la \bar dd\ra^{1/3}$(1 GeV)=-$(229\pm 9)$ MeV&0.02\\
$M^2_0=(0.8\pm 0.1)$ GeV$^2$&0.02\\
$\la \alpha_s G^2\ra=(0.07\pm 0.01)$ GeV$^2$&0.01\\
&\\
Total& 0.16\\
&\\
\hline 
\end{tabular}
\end{table}
where one should notice that, within the present accuracy
of the data on $f_{D_s}$, the result comes mainly from the one used to reproduce
the experimental value of $M_D$. The radiative correction increases the value of $\bar
m_c(M_c)$ by 10$\%$, whilst the non-perturbative term tends to decrease its value by a small
amount of about $3\%$, which indicates that the running mass is mainly of the perturbative
origin. 
\begin{figure}[hbt]
\begin{center}
\epsfxsize=6.cm\epsfbox{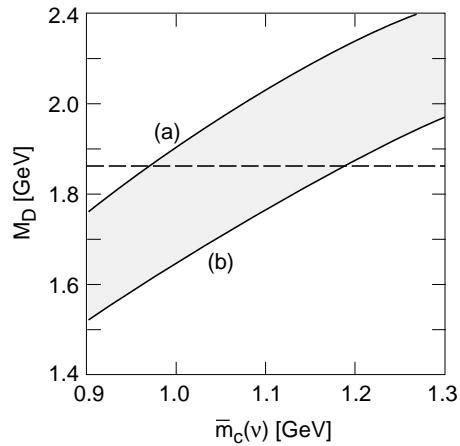}
\caption{Behaviour of $M_D$ versus $\bar m_c(\nu)$. The
horizontal band is the experimental value of $M_D$. The theoretical band is limited by the
two curves: (a) the same as (a) of Fig. 1 but $\nu$=1 GeV and (b)
the same as (b) of Fig. 1 but $\nu$=1.42 GeV.} 
\end{center}
\end{figure}
\nin
We have varied
the subtraction scale $\nu$ in the range from $\tau^{-1/2}\approx$ 1 GeV  to $M_c\approx$ 1.42
GeV, which gives a significant effect (see Table 1) due to the $\log{\nu}$ term appearing
in Eq. (\ref{relation}). Such effect can be eliminated by choosing the subtraction point
at
$\nu=M_c$, at which, we shall extract the results in this paper. 
One should also notice, in Table 1, that the effect of $t_c$ on the result is relatively small from the
value $t_c\simeq 6$ GeV$^2$, where one starts to have a $\tau $ stability until $t_c\geq 7$ GeV$^2$,
where one has $t_c$ stability. The main sources of errors are due to $\Lambda$ and $\nu$.  
Within the errors, the present result is in good
agreement (though less accurate) with the value \cite{SNM}:
\beq
\bar m_c(M_c)=(1.23^{+0.04}_{-0.05})~\mbox{GeV}
\eeq
 obtained, within the same two-loop approximation,
from $M_{J/\psi}$. Inversely, we can use the combined value:
\beq
\bar m_c(M_c)=(1.20\pm 0.05)~\mbox{GeV}
\eeq
 from
$M_D$  and from $M_{J/\psi}$ systems
on the curve $f_D$ as function of $\bar m_c$ given in Fig. 1. Then, one can deduce:
\beq\label{fdrun}
f_{D}\simeq (1.52\pm 0.16)f_\pi,
\eeq
where, as can be seen in Table 1, the errors in this determination come mainly from the
perturbative parameters: $\Lambda$, $m_c$ and
the subtraction scale
$\nu$, and, to a lesser extent, from $t_c$ and the non-perturbative terms. One should notice
that the radiative $\alpha_s$ correction and the non-perturbative terms have
respectively increased the value of $f_D$ by about 25$\%$ and 10$\%$.
We can consider that this result is an improvement of the previous value: 
\beq\label{fdpole}
f_{D}\simeq (1.35\pm 0.07)f_\pi,
\eeq
obtained by using the perturbative pole mass of the charm quark propagator
\cite{SNFD,SNFB}.  However, the good agreement (within the errors)
between the two results in Eqs (\ref{fdrun}) and (\ref{fdpole}) may
be an indirect indication that the pole mass defined at a given
order of perturbation theory, can provide a good description of the
physical process in this channel, and that the  eventual
non-perturbative power corrections induced by the resummation  of
the perturbative series remain small corrections. This observation
can also be supported by the agreement of the value of the
perturbative pole mass obtained here and of the one from the
$J/\psi$ systems, where both values have been obtained at two-loop accuracy. 
A further support
of this argument can also be provided by the agreement of the pole mass deduced 
from Eq. (\ref{relation})
using the value of the running mass, with the one extracted directly in \cite{SNM}. 
Finally,
using the previous value of the ratio
$ f_{D_s}/f_D$ given in Eq. (\ref{fdratio}), one obtains:
\beq
f_{D_s}\simeq (1.75\pm 0.18)f_\pi,
\eeq
which is in a better agreement with the data quoted in Eq. (\ref{fdsex}). A natural
improvement of the analysis done in this paper is
a much more precise measurement of $f_{D_s}$ or/and an evaluation of the QCD 
two-point correlator to
order $\alpha_s^2$, where, to that order, one can have a much better difference
of the perturbative QCD series by the use of either the running or the pole
quark mass.  Both experimental and theoretical projects
are expected to be feasible in the near future.
\section{Extension of the analysis to $f_B$}
We have extended the analysis to the case of the $B$-meson. For this purpose, we use the
two-loop value \cite{SNM,PICH}:
\beq
\bar m_b(M_b)=(4.23\pm 0.05)~\mbox{GeV}
\eeq
from the $\Upsilon$ systems. We obtain the $\tau$ and $t_c$ stabilities respectively at
$0.2-0.3$ GeV$^{-2}$ and $40-50$ GeV$^2$, similar to the case of the pole mass
\cite{SNFD,SNFB}. Then, we
obtain:
\beq
f_{B}\simeq (1.44\pm 0.07)f_\pi,
\eeq
where the errors come mainly and equally from $\Lambda$ and $\bar m_b$. The result is
in good agreement with the analysis where the perturbative pole mass is used:
\beq
f_{B}\simeq (1.49\pm 0.08)f_\pi.
\eeq
This result supports the early '87 sum rule result \cite{SNFB}:
\beq
f_B\approx f_D,
\eeq
indicating a huge $1/m$ correction to the heavy quark symmetry expectation $f_B\sim
1/\sqrt{M_B}$.
\section{Extension of the analysis to $f_{D^*}$}
We have also extended the previous analysis to the case of the vector current having the quantum number
of the $D^*$. In this case, and working with the correlator having the same dimension as $\psi_5(q^2)$
\footnote{Notice that, in this channel, 
the correlator which has less power of $q^2$ than $\psi_5(q^2)$,  and which might present a $\tau$
stability, can have singularities and then should be used with a great care.},
we do not obtain a $\tau$ stability as amonge other things, the coefficients of the chiral
condensates are opposite of the pseudoscalar ones. Then, we conclude that, from this
quantity, we cannot have a good determination of $f_{D^*}$, within our approximation. 

\end{document}